\documentclass[aps, superscriptaddress, twocolumn, floatfix,showpacs]{revtex4}

\usepackage{graphicx}
\usepackage{amsmath, amssymb}
\usepackage[usenames]{color}

\begin{document}

\title{Size-dependent same-material tribocharging in insulating grains}

\author{Scott R. Waitukaitis}
\affiliation{ James Franck Institute and Department of Physics, The University of Chicago, Chicago, IL 60637}

\author{Victor Lee}
\affiliation{ James Franck Institute and Department of Physics, The University of Chicago, Chicago, IL 60637}

\author{James M. Pierson}
\affiliation{Department of Earth and Environmental Sciences, University of Illinois at Chicago, Chicago, Illinois, 60607}

\author{Steven L. Forman}
\affiliation{Department of Earth and Environmental Sciences, University of Illinois at Chicago, Chicago, Illinois, 60607}

\author{Heinrich M. Jaeger}
\affiliation{ James Franck Institute and Department of Physics, The University of Chicago, Chicago, IL 60637}

\date{ \today}

\begin{abstract}

Observations of flowing granular matter have suggested that same-material tribocharging depends on particle size, typically rendering large grains positive and small ones negative.  Models assuming the transfer of trapped electrons can account for this trend, but have not been validated.  Tracking individual grains in an electric field, we show quantitatively that charge is transferred based on size between materially identical grains.  However, the surface density of trapped electrons, measured independently by thermoluminescence techniques, is orders of magnitude too small to account for the scale of charge transferred. This reveals that trapped electrons are not a necessary ingredient for same-material tribocharging.

\end{abstract}


\maketitle

Although tribocharging is typically assumed to arise from frictional contact between dissimilar materials, it can also be caused by interaction between objects made of the same material \cite{Gill:1948, Shinbrot:2008vz}.  Several observations indicate that the mechanism for same-material tribocharging in granular systems is related to particle size, with larger grains typically charging positively and smaller ones negatively.  The electric field of dust devils, for example, is known to point upward, consistent with smaller, negatively charged grains being lifted higher into the air \cite{Leeman:2009iq}.  A similar mechanism is suspected to be responsible for the large electric fields and consequent lightning generated in volcanic ash clouds  \cite{Brook:1974tw, Anderson:1965um, Mather:2006un, Houghton:2013, Cimarelli:2013}.  Zhao {\it et al.}~showed that the charge-to-mass ratio for a variety of powder samples crossed from negative to positive as the particle diameter increased, indicating a similar trend \cite{Zhao:2002tl}.  More recently, Forward {\it et al.}~conducted experiments which revealed a correlation between charge polarity and grain size for samples with a binary particle size distribution (PSD) \cite{Forward:2009kw,Forward:2009in,Forward:2009bv,Forward:2009uq}.

Lowell and Truscott showed that  dragging an insulating sphere across a plane made of the same material usually caused the sphere to charge negatively \cite{Lowell:1986ty}.  They developed a model based on a combination of asymmetry between two contacting surfaces and the transfer of trapped electrons \cite{Lowell:1986ty, Lowell:1986vp}, which they suggested tunnel between surfaces when contact offers the possibility for relaxing into an empty, lower energy state.   If the initial surface density of trapped electrons is uniform, continually rubbing some small region of contact (such as the tip of sphere) across a larger region ({\it e.g.}~a plate) leads to net transfer of charge to the smaller region.  Lacks and coworkers later showed how the same geometrical asymmetry also arises with random collisions among particles of different size \cite{Lacks:2008hz,Lacks:2007un,Duff:2008wk}.  However, while in most situations the transferred charge species is negative, there are some materials, such as nylon, where the  polarity  is reversed, which points to the possibility that other charge species might be responsible (Hu {\it et al.}~recently suggested trapped holes might explain the polarity reversal~\cite{Hu:2012jf}). Given these observations and the lack of quantitative data specifically linking charge transfer to the presence of trapped electrons, their role in same-material tribocharging is uncertain.

Here we test whether or not trapped electrons are necessary for same-material tribocharging.  First, we develop a non-invasive experimental technique that allows us to measure the charge of individual grains while simultaneously differentiating them by size.  For a binary-sized sample, we show that charge is indeed transferred between the different sizes, with large grains becoming more positively charged and small ones more negatively charged.  Assuming the trapped electron model is correct, the amount of charge transferred allows us to put a lower bound on the required trapped electron surface density before mixing.  To test this assumption, we then directly measure the density of trapped electrons on the material surface prior to mixing with a thermoluminescence (TL) technique.  This data puts an upper bound on the actual surface density of trapped electrons that is orders of magnitude smaller than the lower bound required by the trapped electron model.  This demonstrates that trapped electrons are not necessary for same-material tribocharging and suggests that other candidate charge carriers and mechanisms should be considered.

\begin{figure}
\includegraphics[scale=1.0]{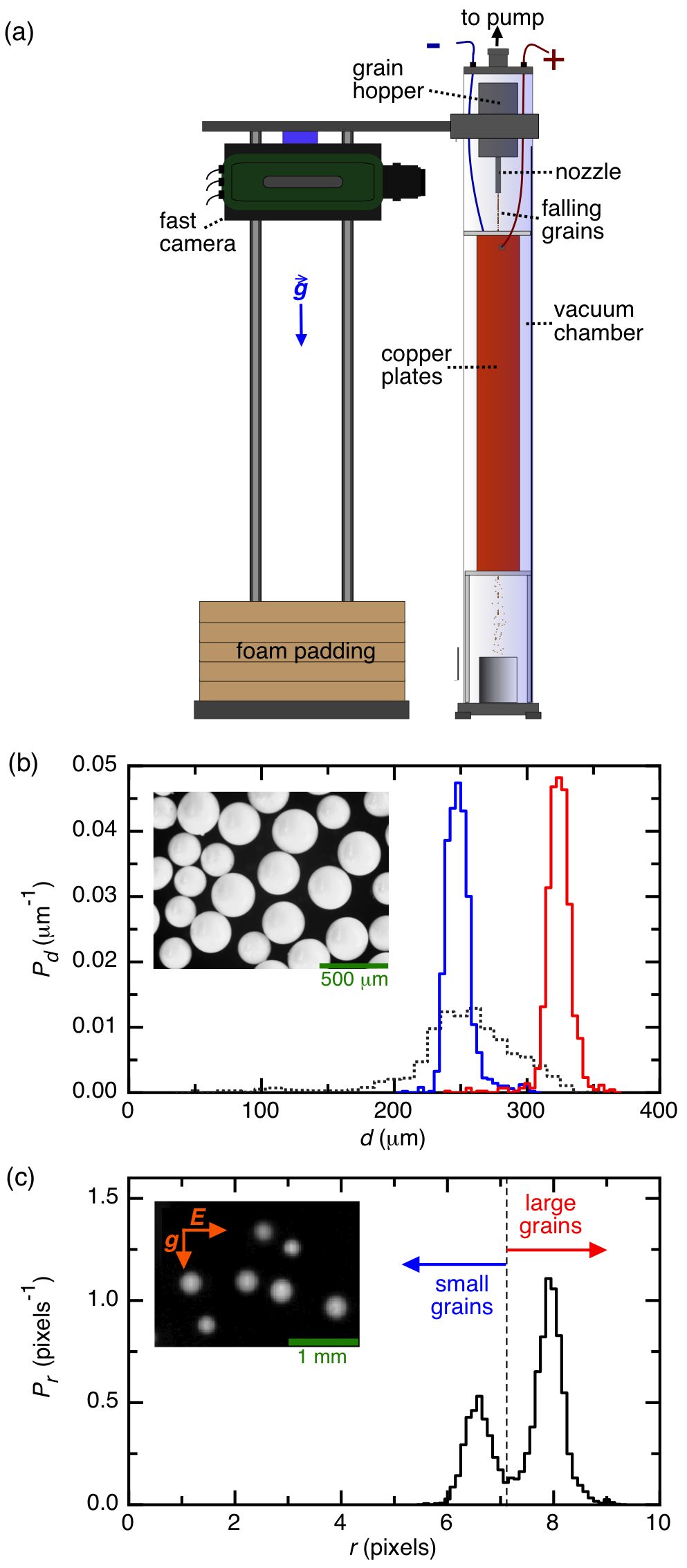}
\caption{(color online).  (a)  Schematic of free-fall charge measurements.  (b) Normalized particle size distribution  determined by optical microscopy for unsifted grains (dotted grey line), sifted ``small'' grains (solid blue line), and sifted ``large'' grains (dashed red line).  Inset:  microscope image of small and large grains.  (c) Radius distribution (pixels) of all grains as determined by analysis of free-fall video (we measure here the ``radius of gyration''--see Ref.~\cite{Crocker:1996} for details).  Dashed vertical line indicates cutoff between ``large'' and ``small'' grains.  Inset: small portion of an image from the high-speed video.}
 \label{fig:size_measurements}
\end{figure}

Our apparatus for measuring individual grain charges while simultaneously differentiating grains by size is shown in Fig.~\ref{fig:size_measurements}(a).  [Here we only discuss the essential details of the measurement technique.  For a full discussion, see Ref.~\cite{Waitukaitis:2013fw}.]  For the granular material, we use fused zirconium dioxide - silicate (ZrO$_2$:SiO$_2$, Glenn Mills Inc.) because it exhibits strong charging behavior and because it is known to the thermoluminescence community for its capacity to store trapped electrons \cite{Iacconi:1978vz, hsieh1994uv,hristov2010uv,azorin1999ultraviolet}.  To ensure the grains are as materially identical as possible, we begin with an initially broad size distribution of grains from a single factory batch.  (We have further confirmed that there is no difference in the composition of the grains with energy dispersive X-ray spectroscopy.)  We take this initial batch and mechanically sieve it into tighter distributions.  We choose two cuts at the tails of the original distribution, the ``large'' and ``small'' grains, and measure their average diameters with an optical microscope, as in Fig.~\ref{fig:size_measurements}(b).  For the experiments here, $\bar{d}_l=326\pm10$ $\mu$m and $\bar{d}_s=251\pm10$ $\mu$m.  We use a Faraday cup \cite{SM} to do a baseline measurement of the mean charges of the large and small grains before mixing, which gives $\bar{q}_l=-(3.1\pm0.3)\times10^4$ e and $\bar{q}_s=-(5.9\pm0.7)\times10^4$ e per grain (here we take ``e'' to be the magnitude of the elementary charge, $+1.6\times10^{-19}$ C).  We then mix the two sizes by fluidizing with air in the grain-coated hopper for approximately 30 minutes.  At this point we put the hopper into the vacuum chamber, as indicated in Fig.~\ref{fig:size_measurements}(a).  Opening an orifice in the nozzle at the bottom of the hopper allows the grains to fall freely via gravity between two large copper plates held at potential difference $V$.  The resulting electric field causes a grain of charge $q$ and mass $m$ to experience a horizontal acceleration $a=qV/ml$.  Outside the chamber a high-speed, high-resolution video camera (Phantom v9.1, 1000 frames per second) guided by low-friction rails falls alongside the grains, which enables us to track their horizontal trajectories with precision and fit with parabolas to extract the accelerations $a$.  The magnification and depth of field of our setup is high enough to allow us to distinguish a particle as ``large'' or ``small'', as shown in Fig.~\ref{fig:size_measurements}(c).  Performing approximately 25 camera drops at a given $V$ allows us to measure the acceleration of several thousand grains and construct independent acceleration distributions for the large and small grains.

\begin{figure}
\includegraphics[scale=1.0]{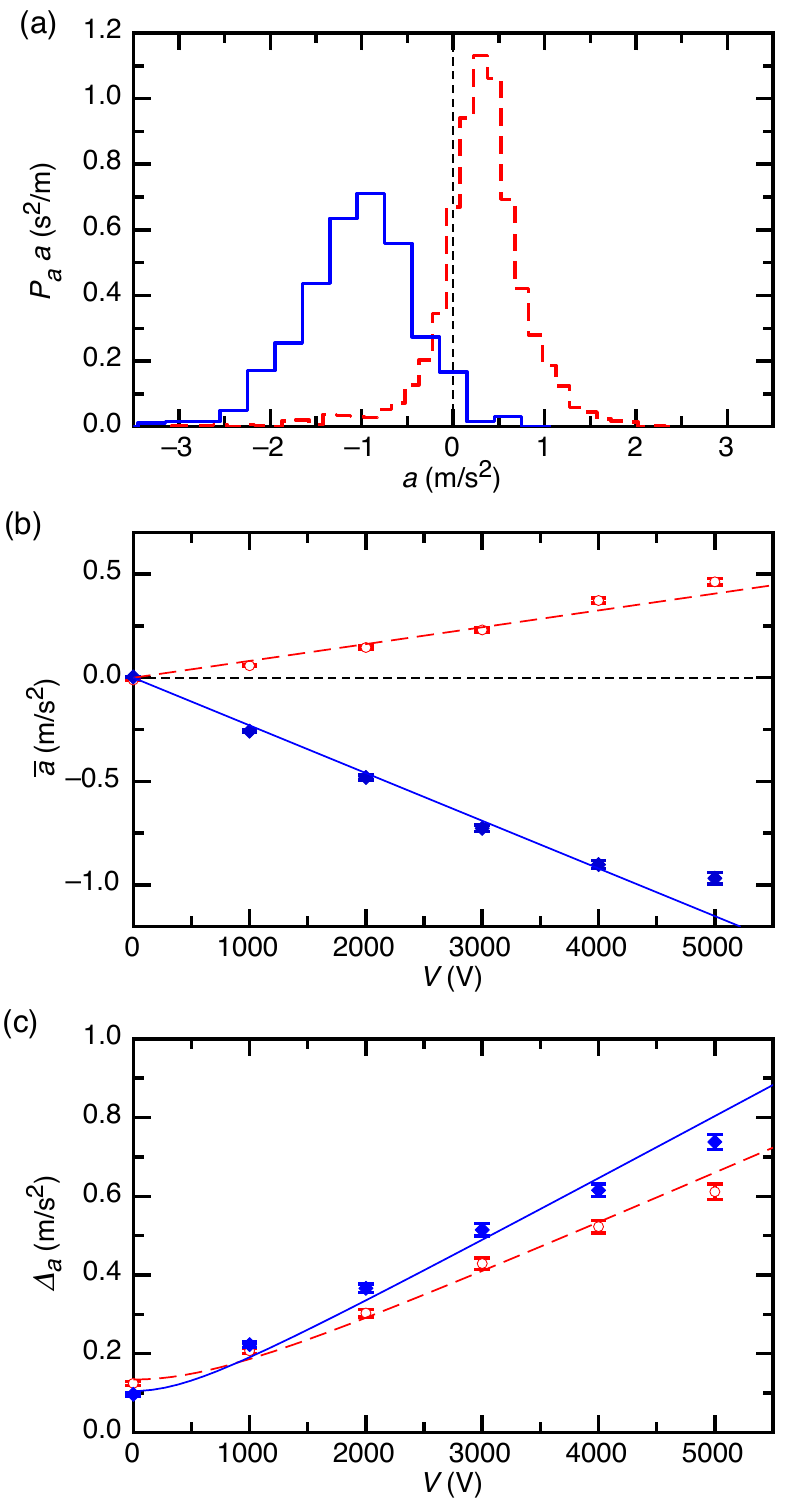}
\caption{(color online).  Size-dependent charging.  (a)  Acceleration distribution of small (solid blue line) and large (dashed red line) grains for $V=3000$ V.  (b)  Mean acceleration $\bar{a}$ of small (blue solid diamonds) and large (red open circles) grains vs.~$V$.  Fits are of the form $\bar{a}=sV$.  (c)  Width of acceleration distributions $\Delta_a$ for small and large gains vs.~$V$ with symbols same as (b).  Fits are of the form $\Delta_a=\sqrt{\delta_0^2 + (kV)^2}$. }
 \label{fig:charge_transfer}
\end{figure}

In Fig.~\ref{fig:charge_transfer}(a), we plot the acceleration distributions for the large and small grains at $V=3.0$ kV ($|E|=59$ kV/m), which shows that the small grains have predominantly negative accelerations, {\it i.e.}~negative charge, while the large grains generally have positive accelerations.  To extract the average charges $\bar{q}_l$ and $\bar{q}_s$, we calculate the mean accelerations $\bar{a}_l$ and $\bar{a}_s$ for each size and plot them as a function of $V$, as in Fig.~\ref{fig:charge_transfer}(b).  The proportionality between $\bar{a}$ and $V$ confirms that the  charge distribution is unaffected by the field  and thus reflects the state of the sample as it exits the hopper (this proportionality would break down if particles collided and transferred charge inside the electric field, as in the mechanism proposed by P\"ahtz {\it et al.}~\cite{Pahtz:2010cq}). From $\bar{a}= s V$, the slope $s=\bar{q}/l \bar{m}$ then gives access to the mean grain charge if the mass is known.   Similarly, the width of the acceleration distribution, $\Delta_a$, is related the the width of the charge distribution,$\Delta_q$, via  $\Delta_a = \sqrt{\delta_a^2+(kV)^2}$, where  $k=\Delta_q/l\bar{m}$ and $\delta_a$ is the average uncertainty in an individual acceleration measurement, independent of applied field.

From the specific material density $\rho$ = 3800 kg/m$^3$ and the PSD in Fig. 1b we compute the average grain masses as $\bar{m}_l=(7.0\pm1)\times10^{-8}$ kg and $\bar{m}_s=(3.1\pm0.8)\times10^{-8}$ kg.  Using the fit values for the slope $s$ this leads to mean charges $\bar{q}_l=(1.8\pm0.2)\times10^{6}$ e and $\bar{q}_s=-(2.3\pm0.6)\times10^6$ e for the two particle sizes.  For the widths we obtain $(2.9\pm0.4)\times10^6$ e and $(1.6\pm0.4)\times10^6$ e for the large and small grains, respectively. Note that the values for the mean charge are two orders of magnitude larger than the residual grain charge prior to mixing.  Within our experimental uncertainties total charge is conserved, which makes it explicit that the charge transfer is occurring among the grains themselves and not with some other material.

Assuming the trapped electron model is correct, the scale of charge transfer between the large and small grains allows us to put a lower bound on the surface density, $\sigma$, of trapped electrons that must have been present before the two sizes were mixed.  If $\sigma$ is the same for all grains initially and {\it all} the excess trapped electrons of the large grains are transferred to the small grains, it must be the case that $\sigma > N/[\pi (\bar{d}_l^2-\bar{d}_s^2)]$, where $N$ is the total number of electrons transferred.  Given the measured number of transferred charges $N\approx2.0 \times 10^{6}$, this implies $\sigma >15$ $\mu$m$^{-2}$.  The randomness of collisions makes this ``complete transfer'' scenario unlikely and, using the results of Lacks {\it et al.}~\cite{Lacks:2007un}, a more realistic estimate is $\sigma >90$ $\mu$m$^{-2}$.

\begin{figure}
\centering
\includegraphics[scale=1.0]{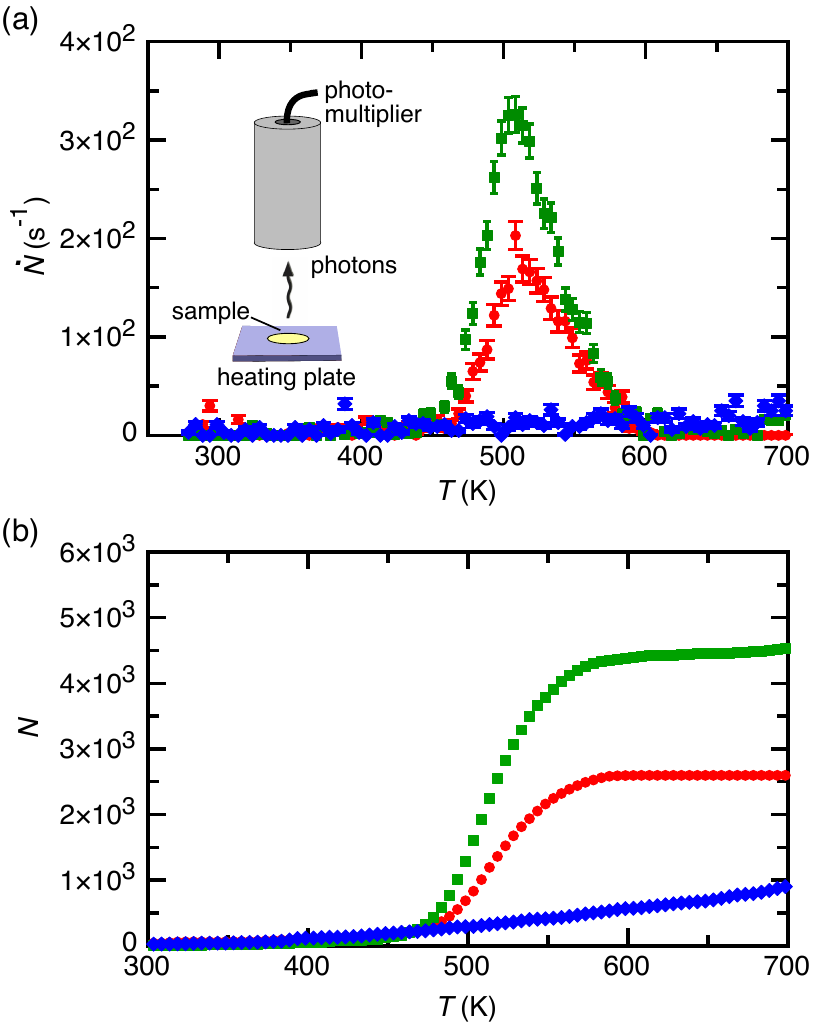}
\caption{ (color online). Thermoluminescence (TL) measurements of trapped state density.  (a)  TL curves of photon count rate $\dot{N}$ vs.~temperature $T$ at ramp rate 6 K/s for untreated grains (blue diamonds), grains exposed to $\sim$12 hours sunlight (red circles), and grains exposed to $\sim$12 hours UV lamp (green squares).  Inset:  schematic of TL measurement. (b)  Total number of photons counted for same data as in (a). }
 \label{fig:thermoluminescence}
\end{figure}

To see if enough trapped electrons to account for the observed charge transfer were present on the pre-mixed grains, we use a technique from thermoluminescence dating.  This is accomplished by heating a sample of the grains with a temperature ramp $T=T_0+\beta t$ while simultaneously measuring the photon emission rate $\dot{N}$ with a photomultiplier [inset to Fig.~\ref{fig:thermoluminescence}(a)].  If trapped electrons are present, one observes peaks in $\dot{N}$ vs.~$T$ because although the emission rate increases with $T$, the available population $N$ in the trap states is being depleted.  [For an introduction to thermoluminescence, we refer the reader to references \cite{Randall:1945uq, Aitken1985, Forman2000}].  In Fig.~\ref{fig:thermoluminescence}(a), we plot typical TL curves taken with a heating rate $\beta=6$ K/s (with a Thorn EMI 9635QB photomultiplier with peak quantum efficiency 0.29 at ~375 nm).  For grains from the same batch as the ones used in the experiments of Fig.~\ref{fig:charge_transfer}, we are unable to detect trapped electrons (the slight rise in the $\dot{N}$ with $T$ is a background ``glow,'' not a TL peak).  If we try to load electrons into the trap states by radiation, either from the sun or from an ultraviolet lamp, we observe one characteristic TL peak.  As explained in \cite{SM}, we can vary the heating rate to show that this trap has an energy below the conduction band $\epsilon=0.36$ eV, typical for the trap depths encountered in other insulators \cite{Aitken1985}.  In Fig.~\ref{fig:thermoluminescence}(b), we plot the integrated number of photons counted for each sample, which shows that even with maximum trap loading no more than $\sim$5000 trapped electrons were present.  Accounting for geometry and the gain of our photomultiplier setup, the actual surface density of trapped electrons is approximately $\sigma= 2\pi N/A_s\Omega$, where $\Omega$ is the solid angle common to the sample ($\sim$5 sr) and the photomultiplier, and $A_s$ is the area of the sample ($\sim$1 cm$^2$). This reveals that the actual density of trapped electrons has an upper bound of $\sigma \approx 1\times10^{-4}$ $\mu$m$^{-2}$, five orders of magnitude lower than the amount necessary to account for the charge transfer we observe in the free-fall experiment.

In principle, it is possible that additional electrons exist at  trap depths deeper  than we can reach with the temperature range available to us, but several factors make this  unlikely.  First, although our TL measurement should be sensitive to traps as deep as $\sim$2 eV (provided the surface density of these traps $\sim$$15-90$ $\mu$m$^{-2}$, as our charging data implies), we see no indication of traps beyond the one at $\epsilon\sim0.36$ eV.  Additionally, traps beyond $\sim$2 eV would be especially deep compared to what is typically reported in the literature \cite{Aitken1985}.  More importantly, if traps did exist in this range, they would be susceptible to unloading via visible light ($\sim1.8-3.1$ eV).  This discrepancy is especially relevant to granular systems continually exposed to visible light from the sun, such as wind-blown dust or volcanic ash, which exhibit strong, same-material tribocharging behavior  \cite{ Freier:1960ta, Leeman:2009iq, Kamra:2007wi, Kok:2008cu, Schmidt:1998ta, Brook:1974tw, Anderson:1965um, Mather:2006un, Houghton:2013, Cimarelli:2013}.  We also considered the implications of possible size-dependent electric discharging, which might occur when the electric field at the surface of a particle exceeds the dielectric strength of the surrounding gas.  However, any discharge would imply that the actual amount of charged transferred between the large and small particles must have been larger than what we measured.  Consequently, the required number of trapped electrons would also have to be larger, which makes the discrepancy with the thermoluminescence experiments more compelling still. 

These considerations lead us to conclude that trapped electrons are not necessary for same-material tribocharging.  This touches on an ongoing debate regarding the most fundamental question in tribocharging:  what is the charge species being transferred?  While in metal-metal tribocharging it has been shown that electrons are transferred \cite{Harper:1951by, Lowell:1975wz}, insulator-insulator experiments have pointed to electron transfer \cite{Liu:2010ip,Liu:2009gb,Liu:2009ga}, transfer of ions adsorbed on the surface \cite{Pence:1994un,McCarty:2008fx,Baytekin:2011bx, Diaz:1998jd}, and recruitment of ions from the atmosphere surrounding the contact \cite{Shinbrot:2008ez}.  

In our case, as the geometric mechanism implies a negatively charged species and trapped electrons are not the culprit, we suspect that ions on the surface or recruited from the surrounding gas might be responsible.  Several recent experiments suggest this could be the case.  Baytekin  {\it et al.}~showed that charge transfer between non-identical insulating materials can be correlated with the breaking of molecular bonds on the surface \cite{Baytekin:2011bx}.  Alternatively, other investigations have pointed out the importance of molecularly thin layers of absorbed water \cite{Pence:1994un, Wiles:2004iu, McCarty:2008fx}.  In particular, McCarty and Whitesides suggest that contact charging between different insulating materials in general might be due to the transfer of OH$^-$ ions.  As they point out, the exact details of how OH$^-$ ions might transfer are not clear, but in this scenario the density of transferrable charges is no longer an issue.  Even with partial monolayer coverage the number of OH$^-$ ions far exceeds the lower bound of 15 $\mu$m$^{-2}$.   Thus, the transfer of OH$^-$ ions in adsorbed surface water is an intriguing possibility that will be the subject of future work.

We thank Gustavo Castillo, Estefania Vidal, Suomi Ponce Heredia, and Alison Koser for contributions during the early stages of setting up the free-fall experiment, Ian Steele for performing the EDS measurements, and  Daniel Lacks, Troy Shinbrot, Ray Cocco, and Ted Knowlton for insightful discussions. This work was supported by the NSF through DMR-1309611.  Access to the shared experimental facilities provided by the NSF-supported Chicago MRSEC (DMR-0820054) is gratefully acknowledged.  S.L.F.~and J.L.P.~acknowledge funding from UIC NSF Grant No.~0850830 and 0602308.  S.R.W.~acknowledges support from a University of Chicago Millikan Fellowship and from Mrs.~Joan Winstein through the Winstein Prize for Instrumentation.

\bibliographystyle{apsrev}

\end{document}